\documentclass[12pt,epsf,epsfig,wrapfig]{article}
\usepackage{graphicx}
\begin{document}
\begin{center}{\bfseries SUMMARY OF THE WORKSHOP DUBNA-SPIN 07 \footnote{Closing Lecture at the Workshop DUBNA-SPIN 07, Sept. 02-09, 2007, Dubna (Russia).}
\vskip 4mm
Jacques Soffer 
\vskip 4mm
{\small {\it Department of Physics, Temple University,\\
Philadelphia, Pennsylvania 19122-6082, USA\\
E-mail: jsoffer@temple.edu}}}
\end{center}
\vskip 4mm
\begin{abstract}During the five days of this workshop we had forty five hours of lectures, so a
tremendous amount of new information was delivered. I will be able only to highlight some aspects of the numerous interesting topics, which were discussed, leaving out many of them.
\end{abstract}
\vskip 5mm 
\section{Introduction}
Let me first begin to say that, given the high density of the scientific program, I had to make a drastic selection and I apologize to the speakers, not or badly, mentioned in this summary. This is partly due to the lack of time and partly to my unability to "`digest"' quickly enough, all this new information. I will not touch technical talks because it is not my field. Fortunately, missing material can be found in these proceedings, collecting all the write-ups of the presentations.\\
From what we heard, it 
is amazing to realize that spin has some relevance all over the places, in a vast energy
range from 100 MeV up to several TeV and in very many different collision processes, namely $e^+e^-$, $e^{\pm}p$, $\mu^{\pm}p$, $\nu p$, $pp$, etc...It is involved in numerous
experiment facilities like, for example, RARF, CLAS, HERMES, HERA, 
COMPASS, BELLE, RHIC, etc....One notices also that significant advances have been achieved recently in polarized beams and targets, allowing to reach higher precision in the new measurements. New projects are under way, which I will just mention: in FAIR at GSI, the PANDA
detector has a broad physics program to study QCD with antiprotons, at Protvino, U70 is
preparing a new polarization program, as well as here in Dubna with the Nuclotron-M.  
On the theory side, the terminology used is also very rich since one has
currently to decode the following sets of initials, PDF, GPD, TMD, DVCS, DIS, SIDIS, DGLAP, BFKL, NLO, NNLO, HT, SSA, etc...\\
Once more, it was clear at this meeting that substantial progress emerge whenever experiment and theory are "`talking to each other"'.
I will try to find the right balence between new experimental results and recent theoretical developments, which have most impressed me, but it was a rather difficult exercice. 

\section{COMPASS Festival}
\begin{figure}[thp]
\begin{center}
\includegraphics[width=85mm]{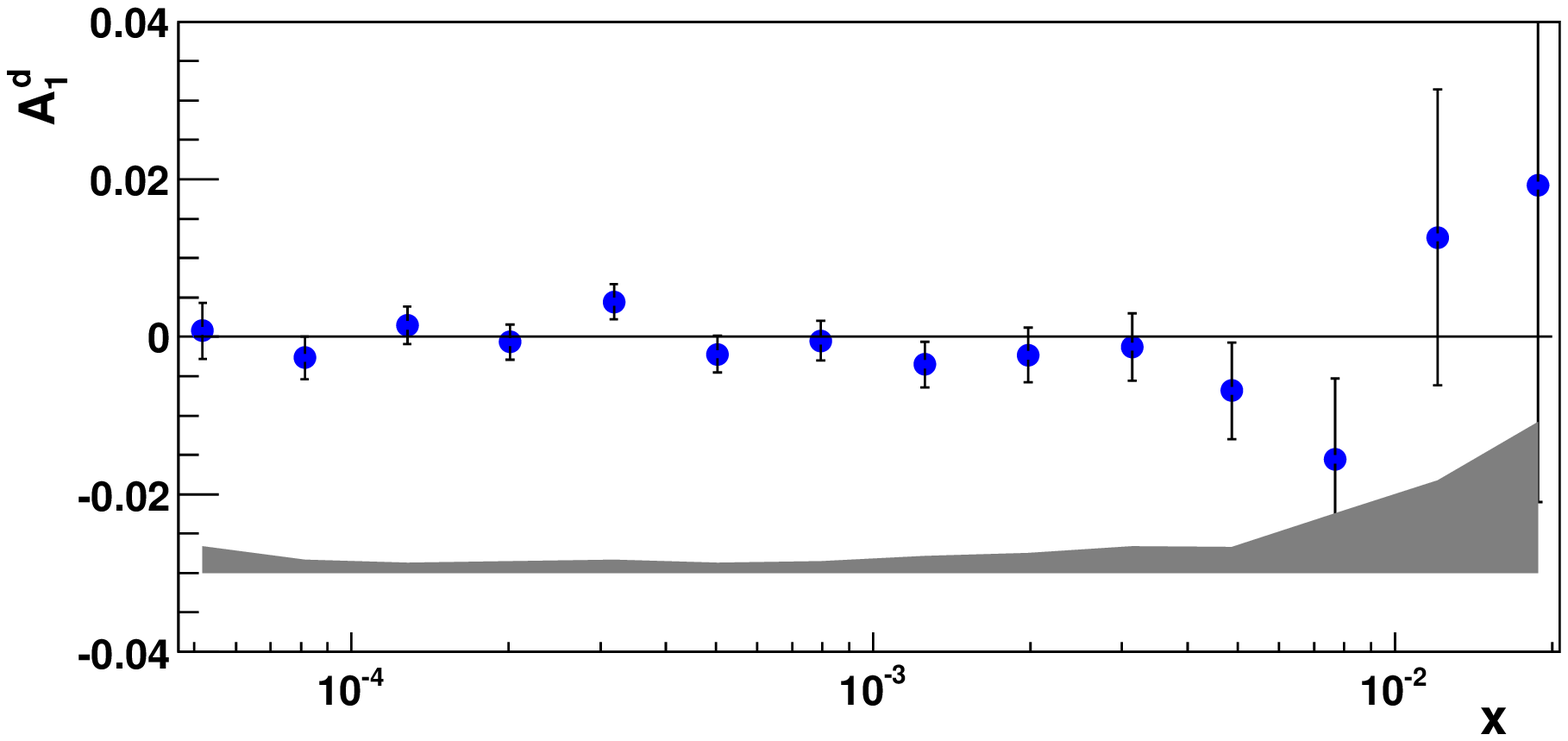}
\includegraphics[width=70mm]{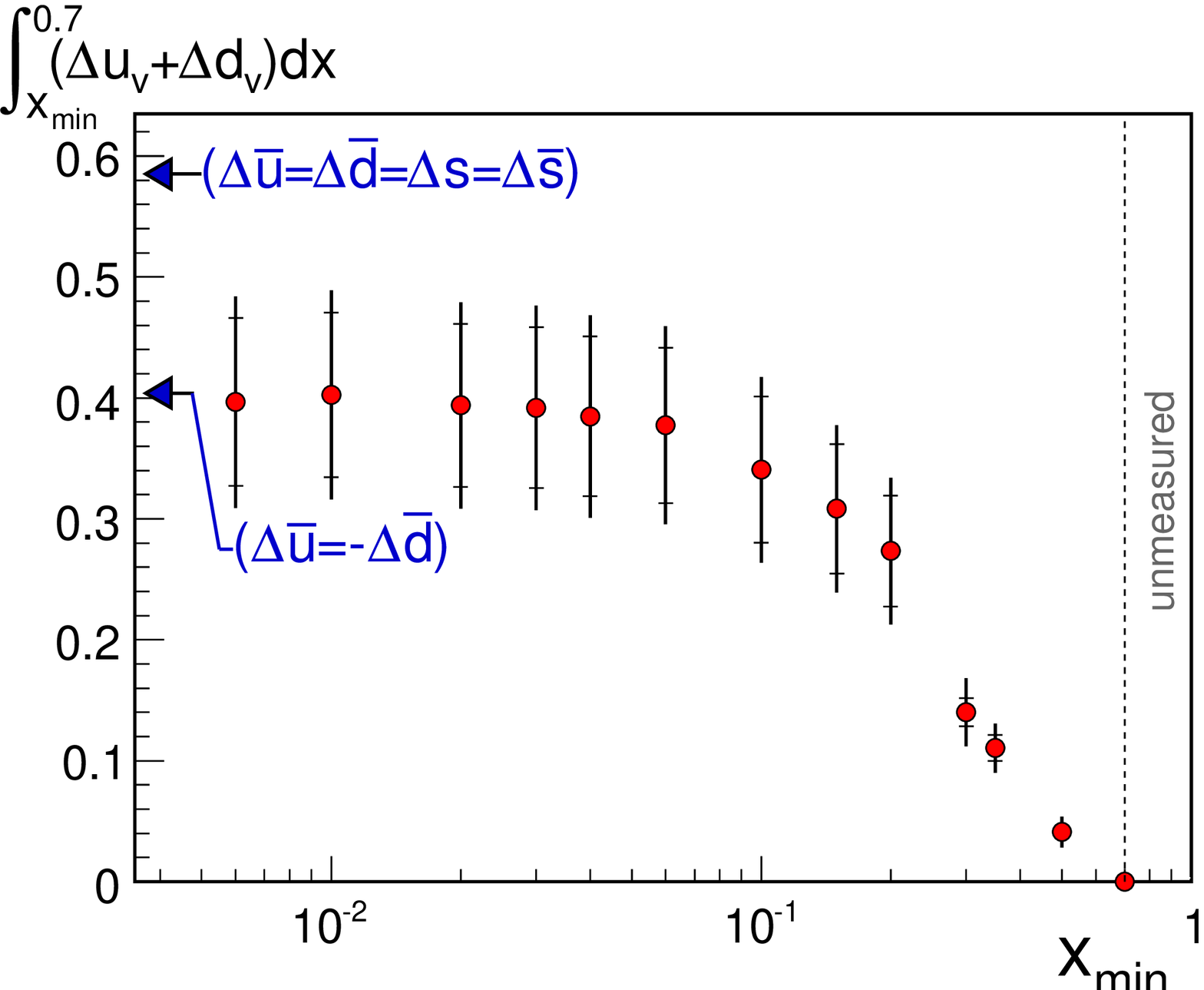}
\caption{\label{fig:1} On the left the asymmetry $A_1^d(x)$ for quasi-real photons ($Q^2 < 1$\, GeV$^2$), as a function of $x$. On the right the integral of $\Delta u_v(x)$$+$$\Delta d_v(x)$ over the range $0.006<x<0.7$, as the function of $x$ minimum, 
  evaluated at $Q^2=10$~GeV$^2$ (Taken from Santos's talk).}
\end{center}
\end{figure}

The COMPASS experiment at the CERN SPS has undertaken a vast experimental
program focused on the nucleon spin structure via
deep-inelastic scattering (DIS) of 160 GeV polarized muons on polarized nucleons. They have obtained very precise results in two kinematic ranges, Q$^2<$ 1 GeV$^2$ and
0.0005 $<x<$ 0.02, as well as 1 $<Q^2<$ 100 GeV$^2$ and 0.004 $<x<$ 0.7, for the spin-dependent structure function $g_{1}^{d}$, by measuring the longitudinal photon-deuteron asymmetry $A_1^d$, with a polarized deuteron target. This asymmetry, shown in Fig.~\ref{fig:1} (Left), is compatible with zero over the small $x$ range and this indicates a strong cancellation between the polarization of the different sea quarks. For large $x$ the asymmetry is large and positive, in agreement with earlier data from SMC and HERMES. They have also discussed the results of a global QCD fit at next-to-leading order (NLO), to the world data on $g_{1}$, which, unfortunately, does
not lead to a unique determination of the gluon polarization $\Delta G$.\\
Another interesting subject is the evaluation of the polarized valence quark 
distributions $\Delta u_v(x)+\Delta d_v(x)$.
The analysis is based on the asymmetry difference A$^{(h^{+} - h^{-})}$,
for hadrons of opposite charges and it gives direct 
access to the valence quark helicity distributions, as 
the fragmentation functions do cancel out. The results, shown in Fig.~\ref{fig:1} (Right), provide information on the contribution of the sea quarks to the nucleon spin. They favour an asymmetric scenario for the sea polarisation, $\Delta {\overline u} = - \Delta {\overline d}$, at a confidence level of two standard deviations, in contrast to the usual 
symmetric assumption, $\Delta {\overline u} = \Delta {\overline d} = \Delta {\overline s} = \Delta s$. However, the statistical errors are still large and do not allow yet a definite conclusion.\\

 The last relevant topic is the gluon polarization $\Delta{G}/G$, which is essential to clarify the spin structure of the nucleon. Since it is impossible to rely on an extraction based on the QCD evolution of the polarized structure functions, COMPASS has chosen to get a direct determination of this quantity, from the measurement of double spin asymmetries in the scattering of polarized muons off a polarized deuteron target.

\begin{figure}[thp]
\begin{center}
\includegraphics[width=8cm]{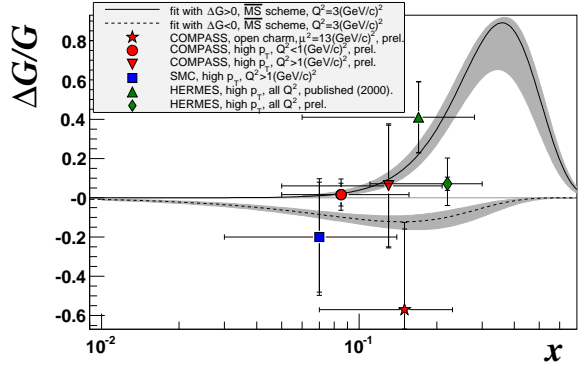}
\caption{\label{fig:2} Comparison of the $\Delta{G}/G$ measurements from various experiments (Taken from Bedfer's talk).}
\end{center}
\end{figure}

 Three different channels sensitive to the gluon distribution are being explored: open charm production and high transverse momentum (high-$p_T$) production, in either the quasi-real (virtuality $Q^2<\,1$~GeV$^2$) photoproduction or the DIS ($Q^2>1$~GeV$^2$) regimes. The first method was described by Y. Bedfer and a preliminary analysis, bearing 2002-2004 data, gives:
\begin{center}
 $\Delta G/G = -0.57 \pm 0.41(stat.) \pm 0.17(syst.)$ ~at $x_g = 0.15 \pm 0.08\ $~and~$\mu^2 = 13 \,GeV^2$.
\end{center}
In his presentation K. Klimaszewski discussed the high-$p_T$ events and reported that the
analysis of combined data from years 2002-2004 leads to a
more precise preliminary result: 
$\Delta G/G = 0.016\pm0.058(stat.)\pm0.055(syst.)$.
The results of COMPASS and from other experiments are shown on Fig.~\ref{fig:2} and they
definitely favor a low value of $\Delta G/G$.

\section{HERMES Festival}
The HERMES experiment at DESY has obtained new results in different area, which were 
introduced in the talk of S. Belostotski. From the analysis of high-$p_T$ 
hadron production, they got the following estimate $\Delta G/G = 0.078\pm0.034(stat.)\pm0.011(syst.)$, with a theoretical uncertainty of $\sim 0.1$.
Polarized inclusive DIS is also used to determine $\Delta \Sigma$, the quark contribution to the nucleon spin, and under some reasonable assumptions, they reported $\Delta \Sigma= 0.330\pm0.025(exp.)\pm0.011(theo.)\pm0.028(evol.)$. Flavor separation for the quark
helicity distributions has been achieved from semi-inclusive DIS data and, in particular,
one gets $(\Delta s + \Delta \bar s) = -0.085\pm0.013(theo.)\pm0.008(exp.)$, by means of $K^{\pm}$ production, which is a preliminary result.

\begin{figure}[thp]
\begin{center}
\includegraphics[width=65mm]{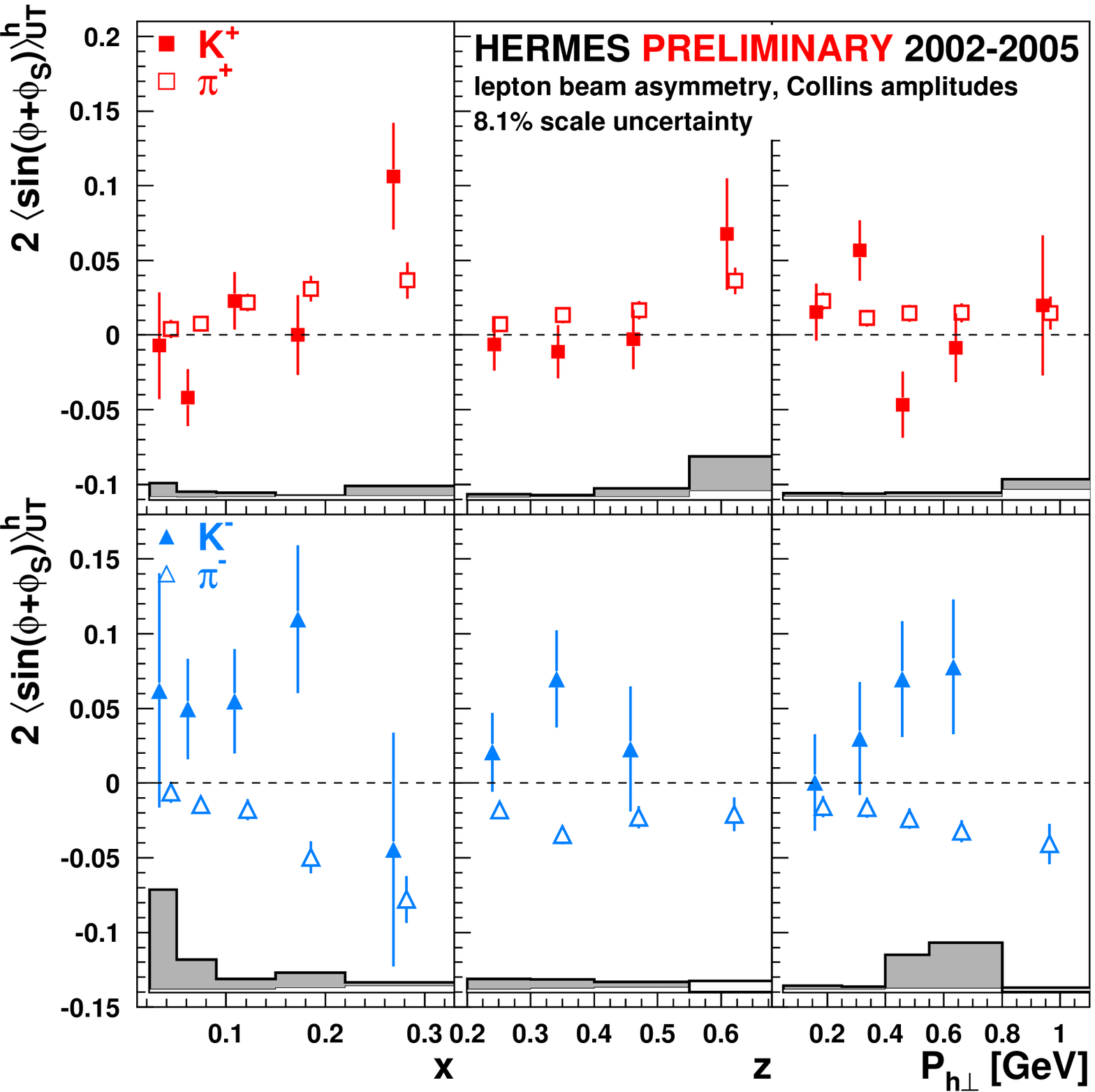}
\includegraphics[width=65mm]{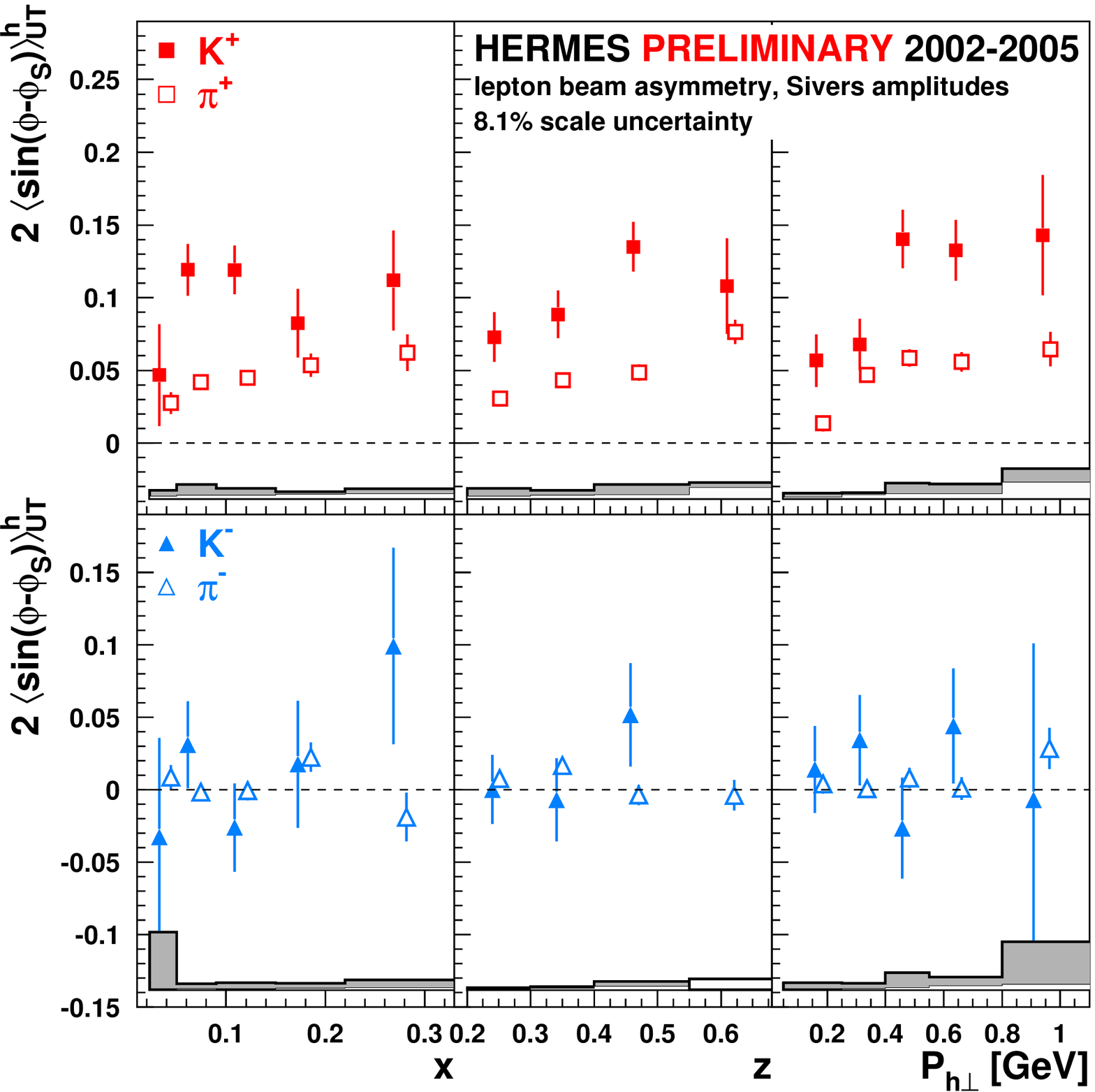}
\caption{\label{fig:3} Collins (left) and Sivers (right) amplitudes for charged pions and kaons,
(as labelled) as a function of $x$, $z$, and $P_{h\perp}$ (Taken from Korotkov's talk).}
\end{center}
\end{figure}

Azimuthal asymmetries were measured in the semi-inclusive production of pions
and kaons and HERMES has collected data with a transversely polarized
hydrogen target from 2002 to 2005. The polarized part of the semi-inclusive 
cross section, for unpolarized beam (U) and a transversely polarized target (T), has 
contributions from both the Collins and Sivers mechanisms. These
asymmetries provide information on the quark Collins and Sivers 
distribution functions. 
These mechanisms produce a different dependence of the azimuthal asymmetry
on the two angles $\phi$ and $\phi_S$, so one can use the variation of $\phi$ and 
$\phi_S$ to disentangle the two contributions experimentally. 
The extracted Collins and Sivers amplitudes for charged pions and kaons, 
are presented in Fig.~\ref{fig:3}, as a function of $x$, $z$,
and $P_{h \perp}$. 
The average Collins amplitude is positive for $\pi^+$ and negative for $\pi^-$.
This is expected if the transversity distribution $h^u_1$ is positive
and $h^d_1$ is negative, like for the helicity distributions.
However, the magnitude of the $\pi^-$ amplitude appears to be as large
as the $\pi^+$ amplitude, which was unexpected. The average Sivers amplitude are significantly positive for $\pi^+$ and $K^+$ and consistent with zero for $\pi^-$ and $K^-$. Note
that the Sivers amplitude for $K^+$ is, by a factor $2.3 \pm 0.3$, higher  
in magnitude than the amplitude for $\pi^+$.

\begin{figure}[thp]
\begin{center}
\includegraphics[width=65mm]{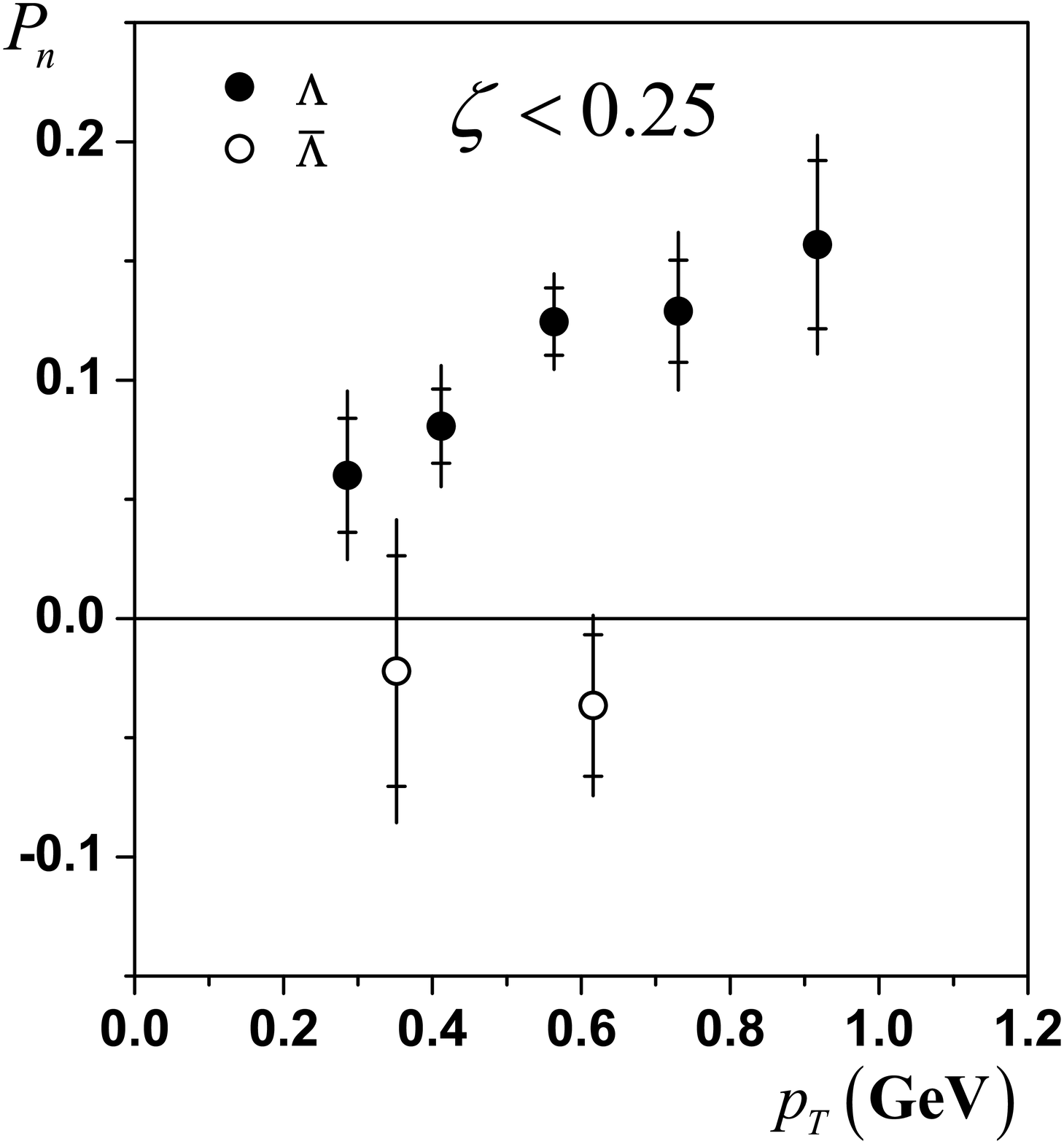}
\includegraphics[width=65mm]{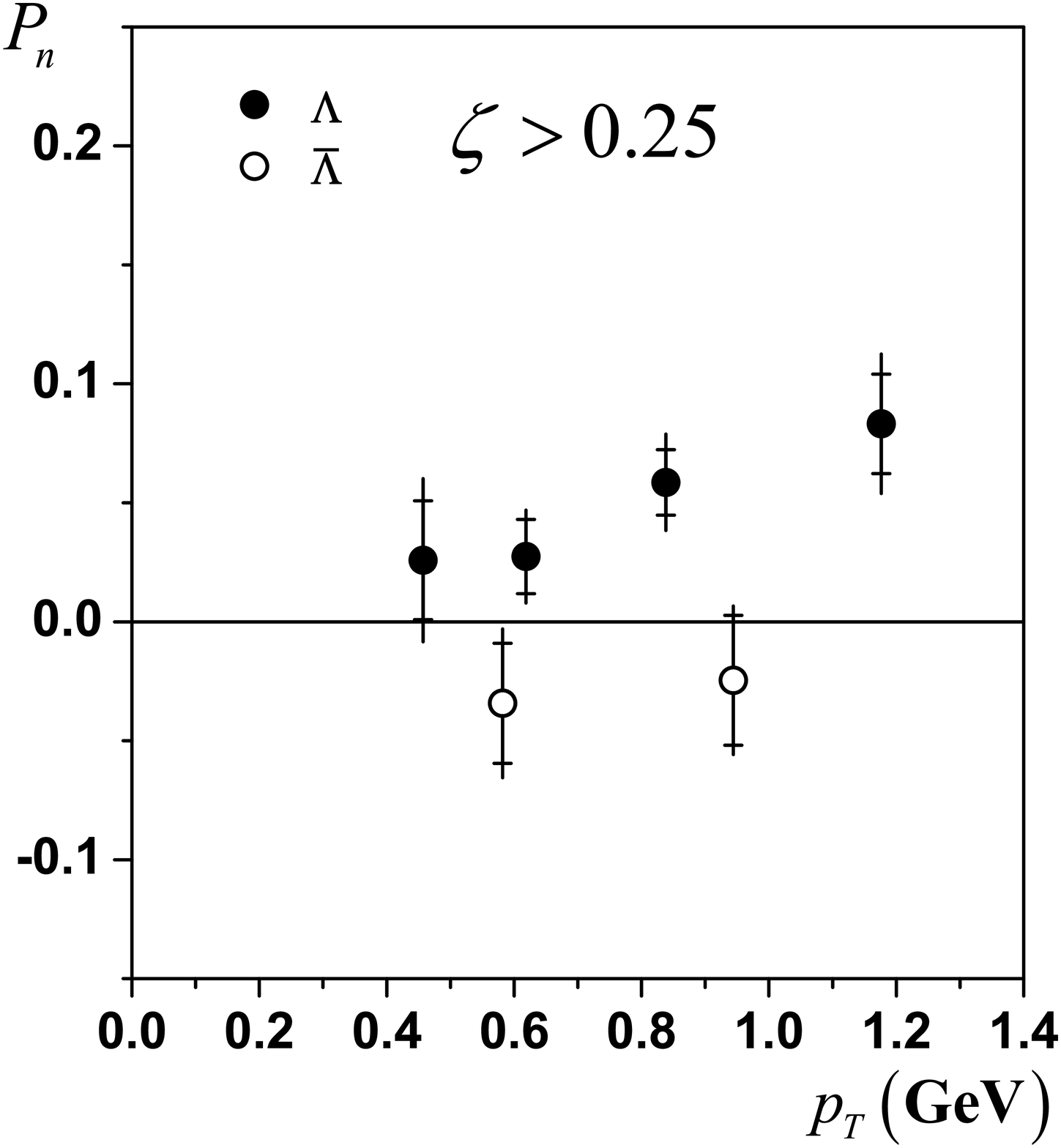}
\caption{\label{fig:4}Transverse polarization $P^\Lambda_n$ and $P^{\bar\Lambda}_n$ as function of $p_T$ for the region $\zeta<0.25$ (left) and $\zeta>0.25$ (right) (Taken from Veretennikov's talk).}
\end{center}
\end{figure}
Transverse $\Lambda$ and $\bar\Lambda$ polarization and spin transfer from longitudinally polarized target have been measured in the HERMES experiment. The kinematic variables are $p_T$ and $\zeta \equiv(E_\Lambda + p^\Lambda_z)/(E_e + p_e)$, where $p_T$ is the transverse momentum with respect to the (lepton) beam, $E_\Lambda$ and $p^\Lambda_z$ are the energy and z-component of the $\Lambda$ momentum (the z-axis is along the lepton beam direction) and $E_e$, $p_e$ are the energy and momentum of the positron beam. In Fig.~\ref{fig:4}, the transverse $\Lambda$ and $\bar\Lambda$ polarizations are shown versus $p_T$ for two kinematical domains $\zeta<0.25$ and $\zeta>0.25$. The $\Lambda$ polarization rises linearly with $p_T$ with higher slope for $\zeta<0.25$ and the $\bar\Lambda$ polarization is consistent with zero.

\section{Belle, BNL and JLab Festival}
We had a very instructive talk by M. Grosse Perdekamp on the analyses of hadronic
events in $e^+e^-$ annihilation at KEK by the Belle Collaboration. He presented the
data on the azimuthal asymmetries between two hadrons produced in the fragmentation of 
a quark-antiquark pair, $e^+ e^- \rightarrow  q \bar{q} \to h_1 h_2 + X$.
The analyses demonstrated that the results on the Collins fragmentation
functions from HERMES and Belle experiments are perfectly compatible. Using these Collins functions the first extraction of the transversity distributions $h^u_1(x)$ and $h^d_1(x)$
was achieved.

\begin{figure}[thp]
\begin{center}
\includegraphics[width=79mm]{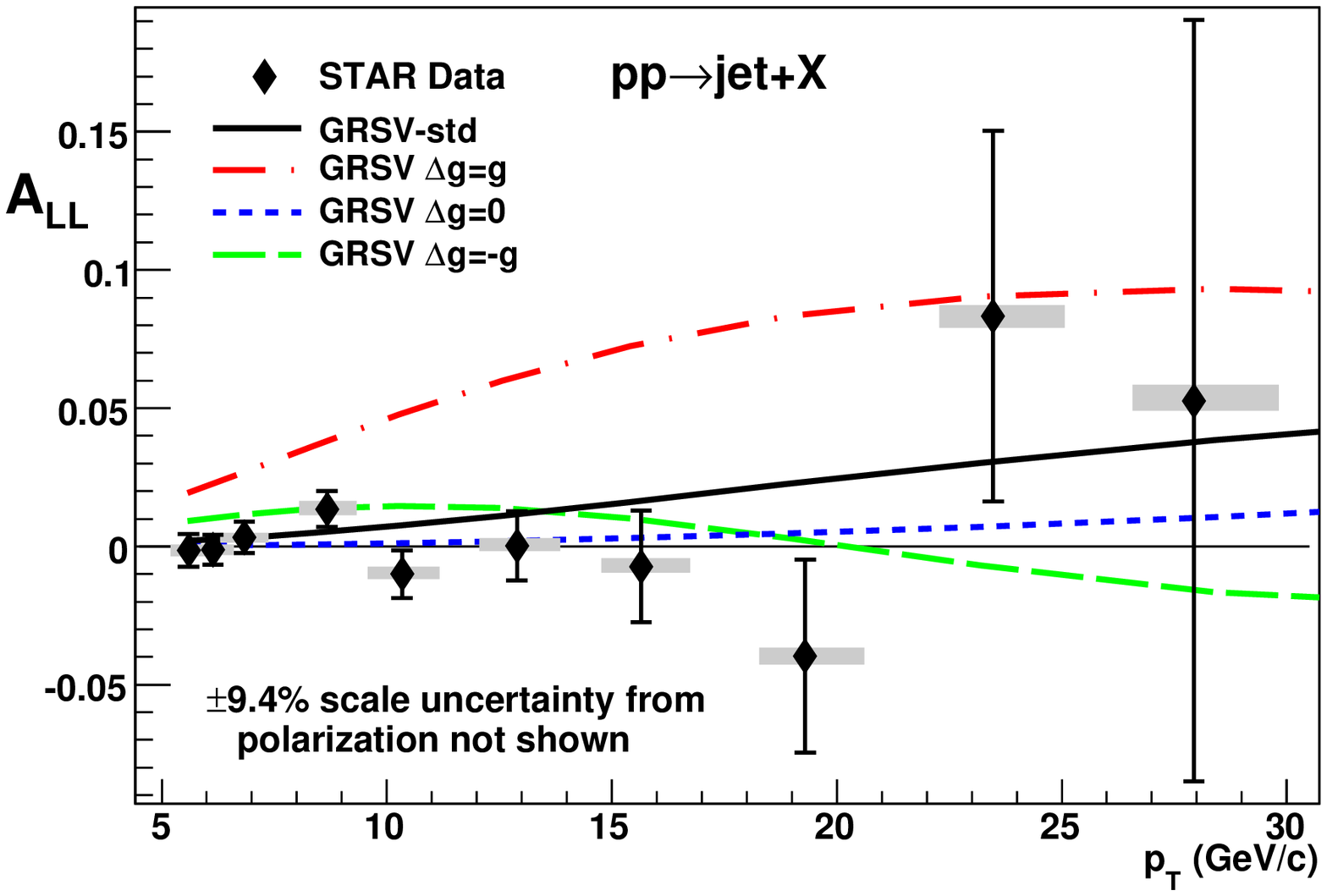}
\includegraphics[width=79mm]{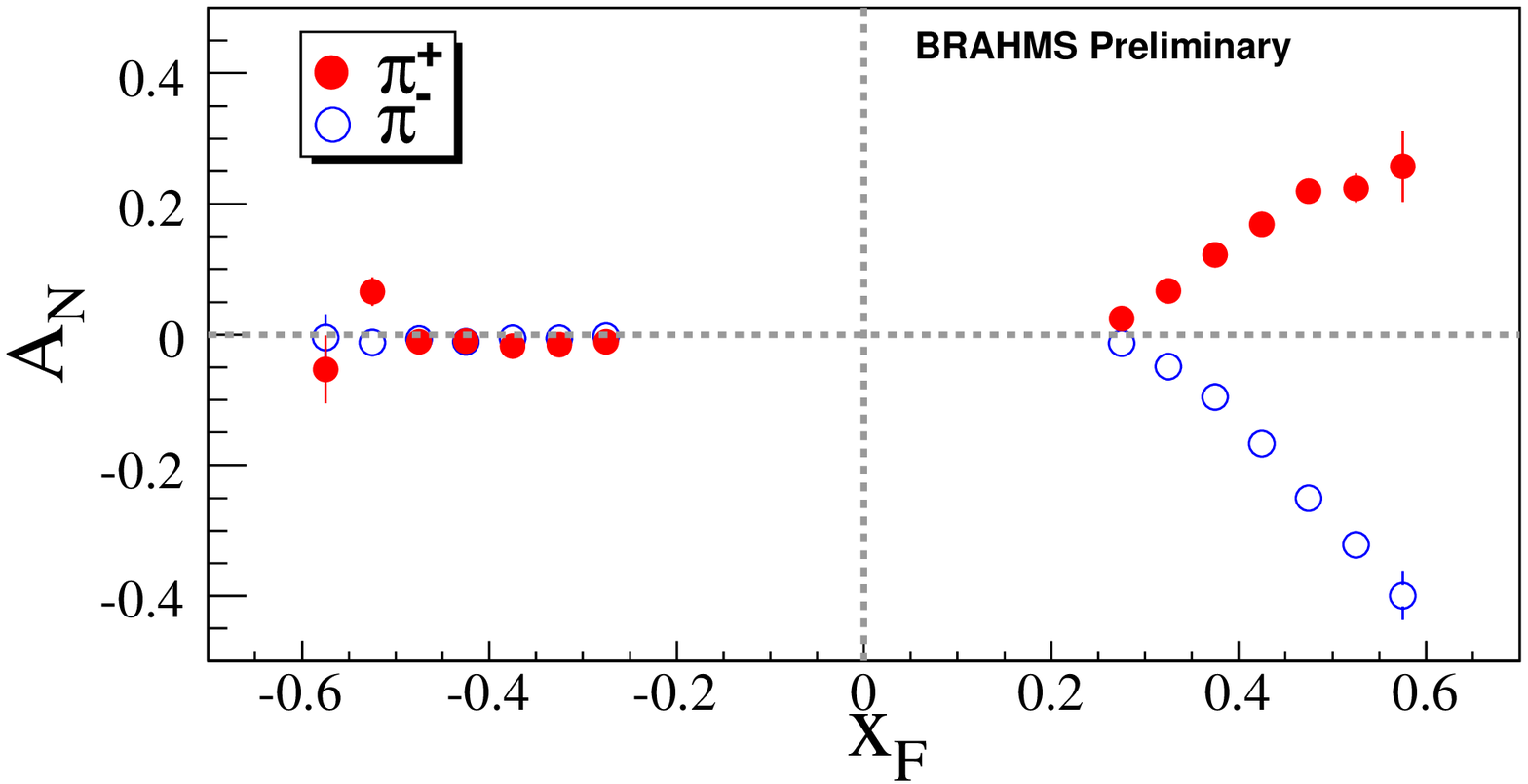}
\caption{\label{fig:5}Left:  Double helicity
 asymmetry $A_{LL}$ for inclusive jet production at $\sqrt{s}$ = 200 GeV versus $p_T$ of the jet from the STAR experiment (Taken from Dunlop's talk). Right: $A_N$ versus $x_F$ for inclusive production of charged pions, at $\sqrt{s}$=62~GeV, preliminary data from 2006,
from the BRAHMS experiment (Taken from Bunce's talk).}
\end{center}
\end{figure}

The RHIC spin program at BNL, underway since 2001, has been presented by G. Bunce.
It consists of colliding polarized protons to study the spin
structure of the proton.  For 2006 they have 
achieved high luminosity collisions at $\sqrt{s}$=200~GeV, with 55 to 60\%
polarization and performed sensitive measurements on the gluon polarization.
Lower $p_T$ production of $\pi^0$ or jets is dominated by the gluon-gluon
graph, and the double helicity asymmetry $A_{LL}$ at mid-rapidity is essentially
quadratic in the gluon polarization.  At higher $p_T$, the quark-gluon
graph dominates, and $A_{LL}$ is linear in the gluon polarization.
The data for $A_{LL}$ for jet production, obtained by the
STAR collaboration, was presented
by J. Dunlop. It is displayed in Fig.~\ref{fig:5} and indicates little or no gluon
polarization in the measured region, which corresponds to
a gluon momentum fraction of $x_{gluon}$ from about 0.02 to 0.3. Fig.~\ref{fig:5} shows also
preliminary results from the BRAHMS experiment for
charged pion transverse spin asymmetries, at $\sqrt{s}$=62~GeV.  The asymmetries at 62~GeV
are very large, and significantly larger than the asymmetries at 200~GeV. At this
energy, STAR has also measured a remarkable asymmetry for $\pi^0$ production, which increases
with $x_F$, for positive $x_F$ and is consistent with zero for negative $x_F$.
A very exciting direction for the transverse spin program is connecting
semi-inclusive DIS and RHIC results. G. Bunce recalled that the final state
interaction needed to generate the asymmetry of DIS and the 
corresponding initial state interaction of Drell-Yan, have different
color interactions, giving in general an attractive force for DIS and a
negative force for Drell-Yan, resulting in opposite sign transverse spin
asymmetries.  This unique prediction of gauge theory must be checked and
this will be done at RHIC.
\begin{figure}[thp]
\begin{center}
\includegraphics[width=45mm]{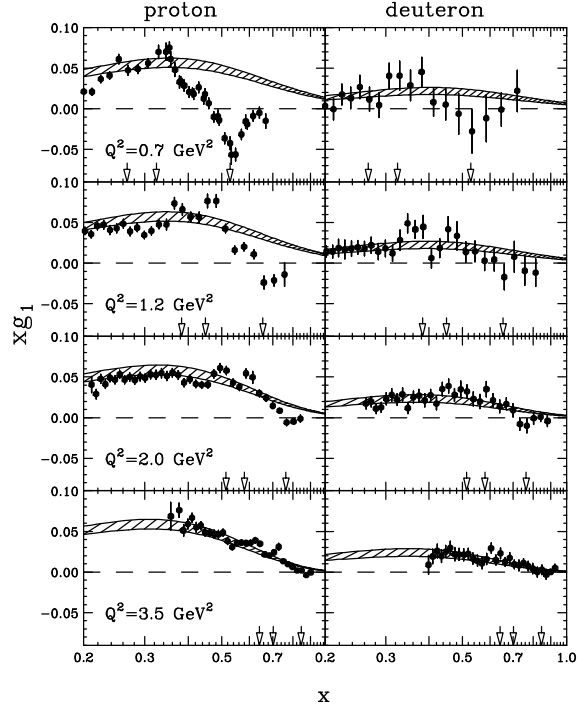}
\caption{\label{fig:6}CLAS data for $xg_1$ in several bins of $Q^2$ for
the proton (left) and deuteron, per nucleon (right) (Taken from Dodge's talk).}
\end{center}
\end{figure}

The CLAS collaboration at Jefferson Lab is pursuing a wide program of
measurements with polarized electrons incident on polarized proton
and deuteron targets, which was partially covered in the talk of G. Dodge. It involves
inclusive, semi-inclusive and exclusive inelastic scattering
over a wide kinematical range in momentum transfer $Q^2$. 
The data are consistent with the expectation that the $A_1$ asymmetry should approach
1 as $x \rightarrow 1$  and they find that
$\Delta d/d$ remains negative up to $x = 0.6$, consistent with results
from Hall A using a $^3$He target. They also studied the onset of quark-hadron duality in spin structure functions.  Quark-hadron duality refers to the observation that the unpolarized
structure function $F_2$, in the resonance region, averages to the smooth
scaling curve for $F_2$ at high $Q^2$.  In Fig.~\ref{fig:6}
one displays $xg_1$ for the proton and deuteron as a function of $x$ for various $Q^2$ bins.  
The high $Q^2$ ``scaling" curve is shown by the hatched
area and indicates the range of $xg_1$ given by PDF fits.  At low $Q^2$ one can see that the data are negative in the region of the $\Delta$(1232) resonance, as expected for a spin 3/2
excitation.  However, as $Q^2$ increases and the $\Delta(1232)$ loses strength, 
the resonances do indeed appear to oscillate about the scaling curve.  
\section{GPD Festival}
\begin{figure}[thp]
\centering
\includegraphics[angle=270, width=0.84\textwidth]{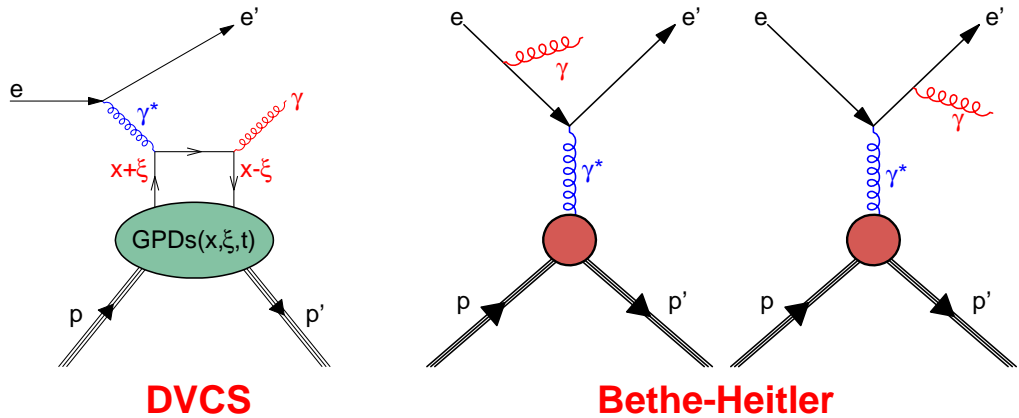}
\caption{ Feynman graphs of DVCS (left) and Bethe-Heitler (right)
processes. Both processes lead to the same final
state, therefore their amplitudes can interfere (Taken from Vilardi's talk).}
\label{fig:7}
\end{figure}

Generalized Parton Distributions (GPD), introduced 10 years ago, is a powerful tool
which offers a way to unify two pictures
of the nucleon, disconnected so far, on the one hand the PDF's
$f(x,Q^2)$, obtained from DIS, and on the other hand the nucleon 
form factors $F(t)$, obtained from $ep$ elastic scattering. The GPD's provide a 
three-dimensional picture of the nucleon and therefore a more 
detailed information on its partonic structure, designated "`nucleon tomography"' 
by N. d'Hose. One hopes to gain some insight
on the localization of partons inside the nucleon and to access to their
orbital angular momentum $L_q$, as first suggested by X. D. Ji. GPD's can be extracted
experimentally through the measurement of hard exclusive reactions,
the cleanest one is the Deeply Virtual Compton Scattering
(DVCS)(or meson production), shown on the left of Fig.~\ref{fig:7}.
In the reaction $e p \rightarrow e' \gamma p$, the Bethe-Heitler (BH) process on the
right of Fig.~\ref{fig:7}, dominates over DVCS in most of the kinematic region.
However, measurable asymmetries in beam spin and beam charge arise
from the interference of both processes. The beam spin asymmetry is
proportional to the imaginary part of the DVCS amplitude, while the
beam charge asymmetry is proportional to the real part of the DVCS
amplitude and both asymmetries can be expressed in terms of GPD's.
  Several models are emerging and predictions made from lattice QCD for the 
first moments of the nucleon GPD's confirm that the transverse size of the nucleon 
depends significantly on the momentum fraction $x$.
The kinematical domain accessible in COMPASS and its availability of positive and 
negative polarized muons gives it a major  
opportunity to measure the different configurations of charge and spin of the
beam, as explained by N. d'Hose. 

Finally, let us mention the results presented by A. Borissov on exclusive diffractive production of light vector mesons ($\rho^0$
and $\phi$) on Hydrogen and Deuterium targets, measured by HERMES in the
kinematic region $0.5 < Q^{2} < 7$ GeV$^2$ and $3.0 < W < 6.3$ GeV.
Data for the $Q^2$ and $W$ dependences of longitudinal cross sections and 
spin density matrix elements are in fair agreement with GPD calculations
based on the `handbag factorization'. This model was presented by S. Goloskokov
and it seems to work well up to HERA energies.
 
\section{Theory Festival}
Several talks were devoted to single spin asymmetries and their connection 
to the Sivers and Collins effects, which generate
the most sizeable single spin asymmetries (SSA)
in semi-inclusive deep-inelastic scattering (SIDIS)
with transverse target polarization, as already mentioned above. 
In his talk A. Efremov gave our present understanding of these phenomena.
Within some uncertainties it was found that the SIDIS data
from HERMES and COMPASS, on the Sivers and Collins SSA from
different targets, are in agreement with each other and with Belle
data on azimuthal correlations in $e^+e^-$-annihilations.
At the present stage of the art, large-$N_c$ predictions 
for the flavour dependence of the Sivers function 
are compatible with data, and provide useful constraints.

The global analysis of HERMES, COMPASS and Belle data reported by A. Prokudin is
leading to the extraction of favoured and unfavoured Collins fragmentation functions and the unknown transversity distributions for $u$ and $d$ quarks,
$h_{1}^{u}(x)$ and $h_{1}^{d}(x)$. They turn out to be opposite in sign, with
$|h_{1}^{d}(x)|$ smaller than $|h_{1}^{u}(x)|$, and both are smaller than their corresponding Soffer bound. This is just a first step for extracting transversities, as noticed by
M. Wakamatsu, who carried out a comparative analysis of the
transversities and the longitudinally polarized PDF's. He concluded that a
complete understanding of the spin dependent fragmentation
mechanism is mandatory for getting more definite knowledge of the
transversities and that some independent determination of transversities is highly desirable,
for example, through double transverse spin asymmetry in
Drell-Yan processes.

O. Teryaev recalled that twist-three quark-gluon correlators were proposed
long ago to explain non-zero SSA and he presented some arguments to establish
a relation between the Sivers function and these twist-three matrix elements.
As a result, the Sivers mechanism may be applied at large momentum transfer. It is
also possible to find some connection between Sivers function and GPD.

D. Sivers discussed chiral dynamics and introduced the concept of spin-directed
momentum transfer from the measurement of a parity-conserving SSA.

In his talk, A. Sidorov  studied the impact of the CLAS and latest COMPASS data on
the polarized parton densities and higher twist (HT) contributions. It
was demonstrated that the inclusion of the low $Q^2$ CLAS data in
the NLO QCD analysis of the world DIS data improves essentially
our knowledge of HT corrections to $g_1$ and does not affect the
central values of PDF's. However the large $Q^2$ COMPASS data
influence mainly the strange quark and gluon polarizations, but
practically do not change the HT corrections. The uncertainties in
the determination of polarized parton densities is significantly
reduced due to both data sets and he concluded  
that it is impossible to describe the very precise CLAS data, if
the HT corrections are not taken into account.

B. Ermolaev presented a description of spin structure function $g_1$ 
at arbitrary $x$ and $Q^2$. It is known that the extrapolation of DGLAP to
the very small-$x$ involves necessarily the singular fits for
the initial parton densities without any theoretical basis. On the
contrary, according to B. Ermolaev, the resummation of the leading logarithms of $x$ is the
straightforward and most natural way to describe $g_1$ at small
$x$. Combining this resummation with the DGLAP results leads to
the expressions for $g_1$ which can be used at large $Q^2$ and
arbitrary $x$, leaving the initial parton densities non-singular.

The talk presented by X. Artru contains two parts. In the first one, he recalls that positivity restrains the allowed domains for pairs or triples of spin observables in polarised reactions, some of which having non-trivial shapes.  
Various domain shapes in reactions of the type ${1/2}+{1/2}\to{1/2}+{1/2}$ are displayed and some methods to determine these domains are mentioned. The second part deals with classical and 
quantum constraints in spin physics, from both discrete symmetries and positivity.

Finally, A. A. Pankov considered the $e^+e^-$ International Linear Collider
(ILC) to study four-fermion contact interactions in fermion pair
production process $e^+e^-\to\bar{f}f$ and he stressed the role played
by the initial state polarization, to increase the potentiality of this future
machine to discover new phenomena.\\

{\bf Acknowledgements}\\

I would like to thank Konrad Klimaszewski for a serious technical help to prepare this talk.
I am grateful to the organizers of DSPIN07, for their invitation to this conference dedicated to L. I. Lapidus, I had the great privilege to meet several times. My special thanks go also to Prof. A.V. Efremov for providing a full financial support and for making, once more, this meeting so successful.\\  
\end{document}